# Wurtzite phase control for self-assisted GaAs nanowires grown by molecular beam epitaxy

*T.Dursap[1], M. Vettori[1], C. Botella[1], P. Regreny[1], N. Blanchard[2], M. Gendry[1], N. Chauvin[3], M. Bugnet[4], A. Danescu[1], J. Penuelas[1*]*

[1] Institut des Nanotechnologies de Lyon – INL, UMR 5270 CNRS, Université de Lyon, Ecole Centrale de Lyon, 36 Avenue Guy de Collongue, F-69134 Ecully cedex, France

[2] Université de Lyon, Université Claude Bernard Lyon 1, CNRS, Institut Lumière Matière, F-69622 Villeurbanne, France

[3] Institut des Nanotechnologies de Lyon – INL, UMR 5270 CNRS, Université de Lyon, INSA de Lyon, 7 Avenue Jean Capelle 69621, Villeurbanne cedex, France

[4] Université de Lyon, INSA de Lyon, Université Claude Bernard Lyon 1, MATEIS, UMR 5510 CNRS, Avenue Jean Capelle, F-69621 Villeurbanne, France

[*] To whom correspondence should be addressed. E-mail : jose.penuelas@ec-lyon.fr

**Abstract**: The accurate control of the crystal phase in III-V semiconductor nanowires (NWs) is an important milestone for device applications. In this work, we present a method to select and maintain the wurtzite (WZ) crystal phase in self-assisted NWs. By choosing a specific regime where the NW growth process is a self-regulated system, the main experimental parameter to select the zinc-blende (ZB) or WZ phase is the V/III flux ratio. The latter can be monitored by changing the As flux, and drives the system toward a stationary regime when the wetting angle of the Ga droplet falls in a target interval, typically in the 90° - 125° range for the WZ phase growth. The analysis of the *in situ* RHEED evolution, high-resolution scanning transmission electron microscopy (HRSTEM), dark field transmission electron microscopy (DF-TEM), and photoluminescence (PL) data all confirm the control of an extended few micrometers long pure WZ segment obtained by MBE growth of self-assisted GaAs NWs with a V/III flux ratio of 4.0.

## Introduction

The occurrence of two crystalline phases in III-V nanowires (NWs) grown using the vapor-liquid-solid (VLS)[1] method has been subject to numerous studies in the last decades. Among significant results, it has been shown that a zinc blende (ZB) or a wurtzite (WZ) phase can nucleate in III-V NWs such as GaAs NWs[2], while only the ZB phase is observed in bulk GaAs. Controlling the crystal phase in such materials would be an important achievement for device applications. WZ and ZB phases exhibit distinct optical[3–5], electronic[6] and piezoelectric[7,8] properties, and a slightly different electronic band structure[9]. The development of a wide range of heterostructures such as quantum dots or monolayer (ML) thin quantum disks and superlattices[10] is also promising.

The ZB phase is mostly obtained in self-assisted GaAs NWs, while the WZ phase mainly occurs in their gold-catalyzed counterparts. Recent studies explained the growth mechanism of the ZB and the WZ phases as depending on the position of the nucleation point of a new ML, either inside the catalyst droplet for the ZB phase or at the triple phase line (TPL) for the WZ phase[2,11–15]. Moreover, it has been shown that the position of the solid nucleus highly depends on the contact angle of the catalyst droplet, and thus on the size of the catalyst droplet[16–19]. Recent experimental studies highlighted a critical contact angle $\beta_{c1}$ in the 121°-124° range for the gold-catalyzed GaAs NWs[16,17], and in the 125°-127° range for the self-assisted ones[18,19], above which the transition from the WZ to ZB phase occurs. Very recently, Panciera *et al.* experimentally observed a second critical angle $\beta_{c2}$ in the 85°-100° range, below which a transition from WZ to ZB phase occurs in self-assisted GaAs NWs[19]. *In situ* transmission electron microscopy (TEM) studies performed by Jacobsson *et al.* highlighted the presence of truncated facets at the top facet of the NWs for wetting angles greater than $\beta_{c1}$, forcing a nucleation inside the catalyst droplet and thus the growth of the ZB phase[16]. The contact angle of the droplet depends directly on its volume, which is mainly function of the growth parameters, and more particularly the Ga and As fluxes in the case of self-assisted GaAs NWs. Thus tuning the Ga and/or the As fluxes to obtain the desired crystal phase, WZ or ZB, has been reported in several studies[12,14,18,20–24].

In order to achieve a precise control on the droplet volume and hence on the contact angle, an *in situ* monitoring technique is highly desirable. Recent studies report on the use of *in situ* TEM to characterize and control the growth of gold-catalyzed GaAs NWs[16,17], and of self-assisted GaAs NWs[19]. However, despite tremendous progress in the development of *in situ* TEM over the last decade, monitoring materials in their real growth environment and growth conditions is still to be achieved for, *e. g.*, the growth of NWs in standard molecular beam epitaxy (MBE) reactors. Reflection high energy electron diffraction (RHEED) on the other hand, is directly coupled to MBE reactors and used to characterize the structural properties during growth. To date only a few studies report on the use of the RHEED for the growth of self-assisted GaAs NWs. For instance, RHEED has only been used to investigate the consumption of the catalyst droplet at the end of the growth[25], while Bastiman *et al.* reported on the incubation time for the growth of the GaAs NWs[26]. The use of RHEED as an *in situ* characterization tool of the growth of NWs coupled with numerical simulations has been reported recently by Jakob *et al.*[27]. Jo *et al.* and Dursap *et al.* have coupled RHEED data with *post mortem* TEM measurements to characterize the crystal phase of catalyst-free InAs NWs[28] and of self-assisted GaAs NWs[29], respectively.

In this work, we achieve the growth of self-assisted GaAs NWs with an extended pure WZ segment, controlled by *in situ* RHEED and based on simulations. We demonstrate the existence of WZ growth conditions to maintain a constant contact angle of the Ga droplet in a desired range. The growth of the WZ crystal phase was confirmed with TEM and PL measurements.

## Conditions for the self-assisted NW growth in the WZ phase

It is well-known that for the self-assisted GaAs growth, the ZB or WZ phase of the NWs is determined by the wetting angle[2,16,17,19], β, defined schematically in the inset of Figure 1. For instance, when the Ga flux is suppressed and the As atoms feed the droplet by direct impingement, the wetting angle of the droplet decreases and the crystal structure evolves from the ZB to WZ phase, and then back to the ZB phase.

Experimental results in Panciera *et al.*[19] show that the WZ phase is obtained for a wetting angle in the (approximate) 90° - 125° range. The main question we address here is the following: is it possible to provide values for the V/III flux ratio that ensure a fixed wetting angle in the 90° - 125° range, thus ensuring the growth of the WZ phase ?

The positive answer to this question is a consequence of the general remark that the self-assisted NW growth is a self-regulated system[30]. Under fixed values of As and Ga fluxes the NW radius, the droplet size, and the wetting angle evolve toward stationary values, a situation in which the amounts of Ga and As atoms feeding the droplet are equal. However, this “constant growth regime” may be attained only once the NW length overcomes a first-stage “transient” regime.

Previous results[20,30–33] on the self-assisted GaAs NW growth show that the Ga atoms are supplied to the droplet through three different sources: diffusion on the $SiO_2$ terminated substrate, diffusion along the NW facets, and direct impingement of the Ga flux on the droplet surface. Meanwhile, the single source of As atoms is the direct impingement on the droplet surface. Recent experimental results obtained by using two Ga sources with different orientations with respect to the substrate[30] show that the contribution from the on-substrate diffusion disappears when the NW length overcomes the diffusion length on the NW facets, which is typically in the 1- 2 $\mu m$ range. For this reason, our experimental procedure starts by a so-called “standard” self-assisted NW growth of 25 min, using a V/III flux ratio equal to 2.4. A Ga growth rate of 0.5 ML/s, quoted in units of equivalent growth rates of GaAs 2D layers measured by RHEED oscillations on a GaAs substrate[34] was used. This procedure then provides NWs longer than 2 $\mu m$. Beyond this length, the Ga supply from the on-substrate diffusion can be neglected, and thus the equal amount of Ga and As atoms supplied to the droplet can be expressed as:

$$S(\alpha_{Ga},\beta,r)q_{Ga} + 2r\lambda_{facet}\sin(\alpha_{Ga})q_{Ga} = S(\alpha_{As},\beta,r)q_{As} \qquad (1)$$

where $q_{Ga}$ (and $q_{As}$) are the nominal fluxes of Ga (and As), $\alpha_{Ga}$ ( and $\alpha_{As}$) are the angles of the Ga source (As source) with respect to the normal to the substrate, $r$ is the NW radius, $\lambda_{facet}$ is the diffusion length of Ga atoms along the NW facets, and $S(\alpha, \beta, r)$ is the projected area of a droplet with wetting angle $\beta$ sitting on top of a NW with radius $r$ on the plane normal to the flux direction, as given by Glas *et al.*[35]. The first terms on both sides of (1) represent the droplet supply by direct impingement, while the second term in the left-hand side (LHS) represents the amount of Ga atoms feeding the droplet by diffusion on the NW facets.

Formula (1) also explains why the system is self-regulated: in the As-rich regime, the right-hand side (RHS) in (1) is larger than the LHS so that the droplet volume decreases. Therefore, the wetting angle decreases also down to a value for which the difference between the direct impingement terms is balanced by the surface diffusion term. In the Ga-rich regime, the volume of the droplet increases, and so does the wetting angle. As both direct impingement terms increase and the V/III flux ratio is > 1, the RHS increases faster than the LHS. In both situations, the evolution of the system tends to a stationary growth regime.

If the source orientations $\alpha_{Ga}$ and $\alpha_{As}$ are known, for a given radius $r$ and diffusion length $\lambda_{facet}$, a wetting angle $\beta \in (90° - 125°)$ can be achieved by using the V/III flux ratio given by:

$$\frac{q_{As}}{q_{Ga}} = \frac{S(\alpha_{Ga},\beta,r)+2r\lambda_{facet}\sin(\alpha_{Ga})}{S(\alpha_{As},\beta,r)} = \frac{S(\alpha_{Ga},\beta,1)+2\gamma\sin(\alpha_{Ga})}{S(\alpha_{As},\beta,1)} \qquad (2)$$

where the last equality holds since $S(\alpha, \beta, r)$ is quadratic with respect to $r$, and $\gamma$ denotes the number $\lambda_{facet}/r$. This equation also shows that a fixed V/III flux ratio leads to different wetting angles for different NW radii, so that the ideal situation is the monodisperse case. Using the model that accounts for the droplet evolution and a variable NW radius presented in Vettori *et al.*[30], the numerical value of $\lambda_{facet} = 1.8\ \mu m$, and our reactor settings ($\alpha_{Ga}$ = 28°, $\alpha_{As}$ = 41°), we compute the values for $\beta(r, \frac{q_{As}}{q_{Ga}})$, further called asymptotic as they equilibrate the amount of Ga and As atoms for various radii and V/III flux ratios.

Figure 1 illustrates the asymptotic values of the $\beta$ angle for a large range of ($q_{As}$, $q_{Ga}$) couples. The red circle in Figure 1 (V/III flux ratio = 2.4 and $q_{Ga}$ = 3.53 atm/(nm$^2$.s)) represents the standard conditions used to initiate the growth process and to overcome the 2 $\mu m$ NW length during the first 25 min of the experiment. As indicated in Figure 1, for a V/III flux ratio greater than 4, the asymptotic value of the wetting angle is lower than 125°. However, larger values of the V/III flux ratio stop the VLS growth process due to the extinction of the droplet.

Scanning electron microscope (SEM) images of a typical sample after a standard 25 min growth is shown in Figure 2a and Figure 2b. The NWs grown were about 3.1± 0.3 *μm* long with a diameter in the 90-100 *nm* range and a density close to 2 *NW/μm²*. The wetting angle was measured to be close to 140°, in accordance with Panciera *et al.*[19] and the expected value circled in red in Figure 1. The typical RHEED pattern obtained along the [1-10] azimuth at the end of the self-assisted GaAs NW growth, *i.e.* after closing the Ga and As shutters, is illustrated in Figure 2c. The visible diffraction spots are uniquely that of the ZB structure.

Thus, the numerical implementation of the model suggests the following recipe: a first VLS growth stage using the standard conditions (ZB phase) in order to overcome the transient regime due to the on-substrate diffusion, followed by an increase of the V/III flux ratio from 2.4 to a value between 4 and 7. We notice that for large diameter NWs the VLS growth process stops, while for NWs with diameters smaller than a critical threshold the VLS growth continues with a wetting angle $\beta > 125°$. However, considering a typical diameter distribution, we estimate that about 90% of the sample will switch from ZB phase to a stationary growth in the WZ phase.

**Experimental evidences of the extended WZ phase**

The stationary growth of the WZ phase was followed in real time using *in situ* RHEED, and the evolution of the RHEED pattern was continuously recorded during the growth of the extended WZ segment. The intensities of the ZB and WZ spots were then extracted in order to obtain the evolution of the RHEED intensity ratios ($\frac{I_{ZB}}{I_{ZB}+I_{WZ}}$) (ZB IR) and ($\frac{I_{WZ}}{I_{ZB}+I_{WZ}}$) (WZ IR) during the growth. Figure 3a illustrates the ZB IR and WZ IR obtained during a 20 min growth with a V/III flux ratio = 4.0, following the 25 min growth with the standard conditions. The horizontal axis represents the growth time under high-As flux. The moment when the As flux increases is designated by t = 0 s. The WZ IR starts to rise around 20 s after the increase of the As flux, as observed in previous studies[29]. A progressive disappearance of the ZB signal during the growth is clearly visible in Figure 3a, as well as in Figure 3b – d, and proves the growth of a WZ phase. The growth was ended by stopping the As and Ga fluxes simultaneously, in order to maintain the droplet at the top of the NWs. The wetting angle of the droplet was then measured on tens of NWs after 5 min, 10 min and 20 min (Figure 3e, f and g, respectively) of growth using SEM. After 5 min and 10 min, the wetting angle was measured close to 120°, as predicted in Figure 1 (circled in green). However, at the end of the 20 min growth, a slight decrease of the diameter is observed, and the wetting angle is around 90°. Thinner NWs were observed and exhibit a bigger droplet at their top, as predicted by the simulations (see SI figure S1). This NW morphology represents about 10% of the NW population and might be responsible of the weak ZB signal observed in Figure 3d. These measurements, coupled to the RHEED IR analysis, are indirect evidences of the growth of an extended WZ segment. The presence of the droplet after a 20 min growth using a V/III flux ratio = 4.0 suggests that the growth of the WZ segment could be extended even further. Additional IR curves and RHEED patterns obtained using different V/III flux ratios are illustrated in Figure S2 of the Supporting Information.

The crystalline structure of these NWs was investigated using high-resolution high angle annular dark field (HAADF) imaging in the scanning transmission electron microscope (STEM), as well as dark-field (DF) TEM imaging. Figure 4a shows a STEM-HAADF overview of a single NW after the 45 min growth (25 min growth with a V/III flux ratio = 2.4 followed by a 20 min growth with a V/III flux ratio = 4.0). The NW initially crystallizes in the ZB structure, as evidenced in the atomic structure revealed by the STEM-HAADF image in Figure 4c. The transition region between the ZB and WZ phase consists of alternate WZ and ZB domains, as highlighted in the DF-TEM image in Figure 4b. An example of such succession of WZ and ZB domains is shown in the STEM-HAADF image in Figure 4d. After the transition, a WZ segment of about 1.3 $\mu m$ long is evidenced in the DF-TEM image, obtained using a $[1\bar{1}00]$ g vector specific to the WZ phase, as shown in Figure 4b.  Only 2 stacking faults (SFs) are visible in the whole segment (pointed by the red arrows in Figure 4b), separating the 1.3 $\mu m$ long WZ segment into 3 sub-segments of 620 *nm*, 180 *nm* and 480 *nm* of pure WZ. The WZ structure is confirmed by high-resolution STEM-HAADF, as shown in Figure 4e, up to the head of the NW (Figure 4f). It is worth noting that the last few crystallized atomic planes, just below the Ga droplet, correspond to the ZB structure. This portion of ZB crystal is expected due to a wetting angle close to the critical angle $\beta_2$, in the 85° – 100° range[19], when the NW growth was purposely stopped. Further investigation of the NW atomic structure by STEM-HAADF reveals that the ZB and WZ segments of the NW are As-polarized, with a [111] growth direction, as shown in Figure 4g and Figure 4h respectively. In the present growth conditions, this result is in agreement with the expected polarity of the ZB and WZ crystalline phase[36].

To investigate the photoluminescence (PL) of the sample, we use a specific property of high refractive index NWs: due to the waveguiding properties of the NWs, the absorption efficiency is strongly related to the incident light wavelength λ[37]. In the case of a 95 nm diameter GaAs NW, the absorption is strongly localized in the upper part of the NW (WZ section) when λ=532 nm, whereas the absorption is mainly observed in the bottom part (ZB section) when λ=671 nm (see Figure S5 in Supporting Information). Figure 5 shows the 12 K PL spectra of the sample for a 532 nm and a 671 nm continuous wave optical excitation at low excitation power (21 $W/cm^2$). Both spectra reveal a broad emission in the 1.46-1.49 eV range. This energy range is known to be related to a type II ZB/WZ emission [38,39], which is consistent with the TEM images. When the sample is excited by a λ=671 nm laser, a peak is observed at 1.516 eV (peak A), i.e. 3 meV below the low temperature band gap of ZB GaAs ($E_{gap}$=1.519 eV). If peak A is related to the recombination of free excitons, their recombination energy should be equal to:

$$E = E_{gap} + E_{strain} + E_{QC} - E_{binding} \tag{3}$$

where $E_{strain}$, $E_{QC}$ and $E_{binding}$ are the energy shifts induced by the strain, the quantum confinement and the exciton binding energy, respectively. Firstly, we can neglect the strain in our NWs: no shell has been grown to passivate the GaAs and the NWs are not lying on a host substrate avoiding any substrate-induced strain during sample cooling [40,41]. The quantum confinement, if any, is weak. Using the formula of Ref [42], a 1-2 meV confinement energy is expected for a cylindrical shaped ZB GaAs NW with a 90-100 nm diameter. Secondly, an exciton binding energy in the order of 4.2 meV has been reported for bulk ZB GaAs [43,44]. Therefore, the peak A emission energy agrees with the recombination of free excitons in ZB GaAs and with a laser absorption in the lower part of the NWs.

In the case of the 532 nm excitation wavelength, the PL emission is quite different: the PL emission is dominated by a peak at 1.5235 eV (peak B), whereas peak A appears as a low-energy shoulder. Peak B, located above the ZB GaAs band gap, is only observed when a WZ segment is grown on the upper part of the NWs (see Supporting Information). As a consequence, we assume that peak B is related to free exciton recombinations in the WZ section of the NWs in agreement with the strong absorption of the 532 nm laser in the upper part of the NWs. This result is also close to low temperature PL studies performed by other groups on WZ GaAs NWs grown by MOVPE (catalyst-free) or by MBE (Au or Mn catalysts)[45, 46, 47], where WZ emission is observed at 1.522 eV, 1.519 eV, and 1.518 eV, respectively. To determine the low temperature band gap of WZ GaAs, we must consider the quantum confinement and the exciton binding energy for this crystallographic phase. The calculations are performed using the dielectric constant $\varepsilon_0 = \sqrt{\varepsilon_0^{\perp}\varepsilon_0^{\parallel}} = 12.77$ from ref [48] and an exciton reduced mass μ=0.05-0.06 from ref [49]. We find that the quantum confinement (1-2 meV) and the binding energy (4-5 meV) of WZ GaAs are very similar to those of ZB GaAs. Therefore, we estimate that the band gap of WZ GaAs is located 6-9 meV above that of the ZB phase.

We notice that both peaks are quite narrow: about 7 meV for peak A and 8.5 meV for peak B. These linewidths are comparable or better than the 7 meV [45] and 18 meV [46] values reported for non-passivated WZ GaAs NWs and not far from the 4 meV linewidth obtained on single passivated NWs [47,50].

In conclusion, numerical simulations were used to provide a set of parameters meant to stabilize the wetting angle of the Ga droplet in self-assisted GaAs nanowires grown by MBE for an extended time, thereby obtaining a few $\mu m$ long WZ segment: the applicability of this theoretical model was demonstrated experimentally. The growth of the ZB or WZ crystalline phase was monitored by *in situ* RHEED. Evidence of a 1.3 $\mu m$ long WZ segment, grown with a V/III flux ratio equal to 4.0, was provided by high-resolution STEM-HAADF and DF-TEM analysis, and the luminescence properties were shown by PL measurements. This work demonstrates that a precise tuning of the As flux opens the way to control the length of pure WZ crystalline segments in self-assisted GaAs nanowires. The possibility to grow an extended WZ segment from the beginning of the growth of the self-assisted GaAs NWs could be investigated by a constant tuning of the As flux, so as to compensate the variation of the amount of Ga supplied to the droplet. This combined experimental and numerical demonstration is of major interest to tune the properties of III-V nanostructures on demand, and for the fabrication of semiconducting heterostructures with novel functionalities.

**Methods**

The GaAs NWs were grown on epi-ready Si(111) substrates using a solid-source MBE reactor. Each substrate was cleaned in ultrasonic bath during 5 min in both acetone and ethanol, and degassed at 200°C in ultra-high vacuum before introduction inside the MBE reactor. On each substrate, the 2 *nm* native $SiO_2$ oxide was preserved to enable the self-assisted growth[51]. To form the Ga droplets, the substrate was heated to 450°C and 1 monolayer (ML) of Ga was pre-deposited[52,53] at a deposition rate of 0.5 ML/s, quoted in units of equivalent growth rates of GaAs 2D layers measured by RHEED oscillations on a GaAs substrate[34]. Then, the substrate temperature was increased to 600°C, the growth temperature. Finally, the opening of the Ga and As fluxes initiated the growth of the NWs. The MBE system was handled by a homemade software that finely controls the Ga and $As_4$ fluxes, the valves and shutters. The NWs were grown with Ga and $As_4$ fluxes of 0.5 ML/s and 1.2 ML/s, respectively, corresponding to a V/III flux ratio = 2.4. In order to investigate the influence of the V/III flux ratio on the wetting angle, different As fluxes (1.5 ML/s, 1.75 ML/s, 2.0 ML/s and 2.15 ML/s) were used for the growth of the extended WZ segment and for each As flux, three different growth times were used (5 min, 10 min and 20 min). All the samples were characterized by RHEED at 30 keV to obtain real time information on the crystal structure evolution of the NWs. The samples were rotating during the growth under the standard conditions. The rotation was then stopped during the growth of the extended WZ segment in order to precisely record the intensity evolution of the ZB and WZ spots. The RHEED intensity ratios and RHEED diffraction patterns of each experiment are represented in SI Figure S2. Each sample was then observed and characterized with a JEOL scanning electron microscope (SEM) using an acceleration voltage of 10 kV. The WZ and ZB phases were differentiated at high resolution by scanning transmission electron microscopy (STEM) - high angle annular dark field (HAADF) imaging, in a JEOL JEM-ARM200CF NeoARM, equipped with a cold-FEG, a last generation aberration-corrector (CEOS ASCOR) of the probe-forming lenses, and operated at 200 kV. Dark field TEM images were obtained using a JEOL 2100HT microscope operated at 200 kV. The photoluminescence measurements were performed at 12 K in a closed cycle helium cryostat. Optical excitation was provided by a continuous wave diode-pumped solid-state laser (532 nm or 671 nm wavelength) with a ≈ 200 µm spot size. The PL was collected through a Cassegrain reflector and analyzed using a liquid-nitrogen cooled silicon based array detector coupled to a monochromator.

**Acknowledgements:**

The authors thank the NanoLyon platform for access to the equipments and J. B. Goure for technical assistance. The authors acknowledge the French Agence Nationale de la Recherche (ANR) for funding (project BEEP ANR-18-CE05-0017-01). N. C. acknowledges funding from the ANR (project HEXSIGE ANR-17-CE30-0014-04). The (S)TEM work was performed at the consortium Lyon-St-Etienne de microscopie. The authors are grateful to Y. Lefkir and S. Reynaud for technical assistance using the Jeol NeoARM instrument.

**Figures**

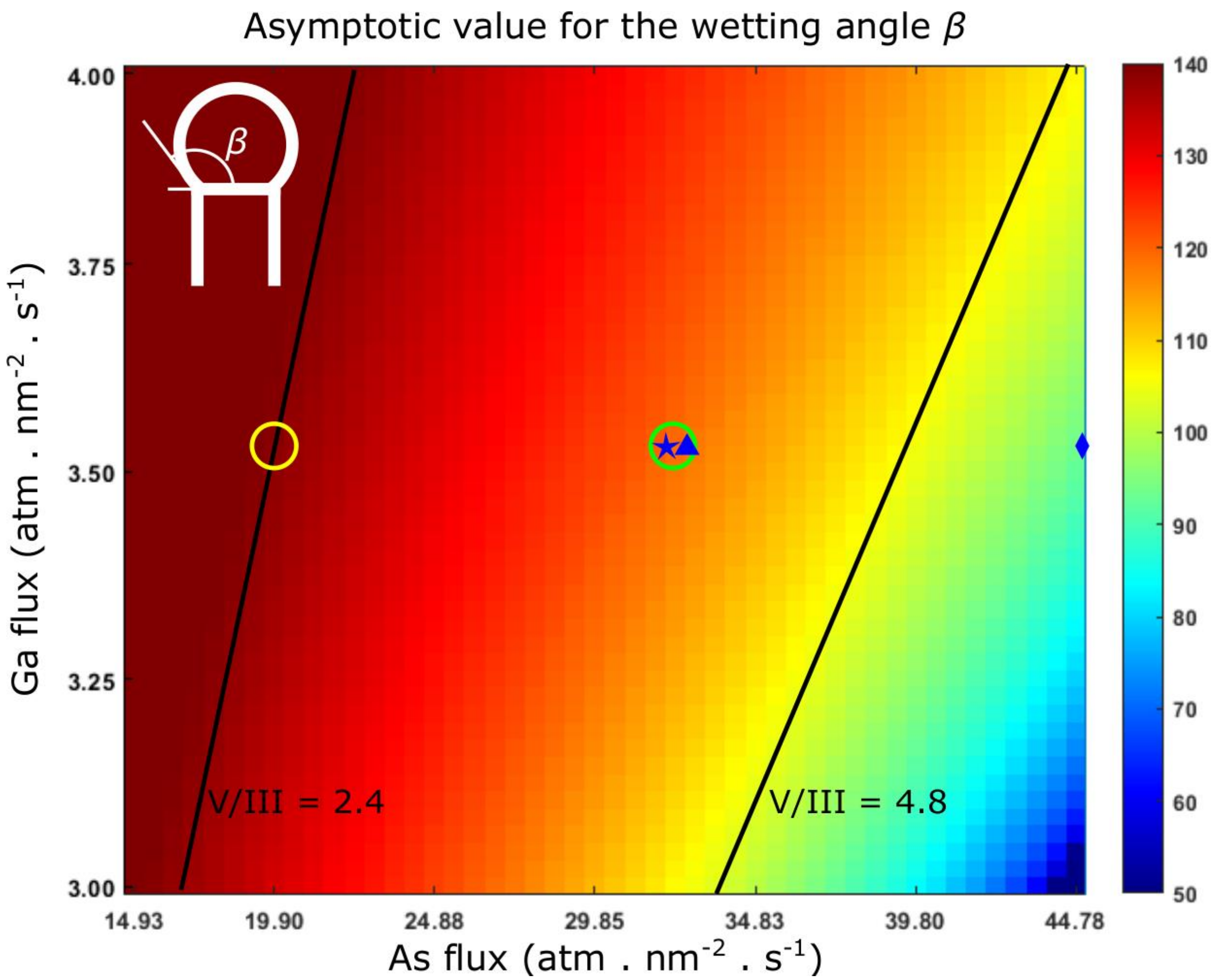

**Figure 1:** Values of the wetting angle $\beta$ obtained for NWs with $r$ = 50 nm for various V/III flux ratio. The area highlighted with the yellow circle represents the conditions for the NW growth (V/III flux ratio = 2.4 at $q_{Ga}$ =3.53 *atm/(nm$^2$.s)*) in the first 25 min, needed in order to overcome the transient regime where diffusion on the $SiO_2$-terminated substrate contributes to the Ga supply. Values lower than 50° are not represented as we consider that the droplet disappears and the VLS growth stops. The green circle represents the expected wetting angle obtained with our extended WZ growth conditions (V/III flux ratio = 4.0 at $q_{Ga}$ = 3.53 *atm/(nm$^2$.s)*). The experimental values of the wetting angle obtained on tens of NWs after an extended WZ growth of 5 min, 10 min and 20 min are represented with the blue star, triangle and diamond, respectively. Additional figures for $r$=40 nm and $r$ =60 nm are presented in the SI figure S1.

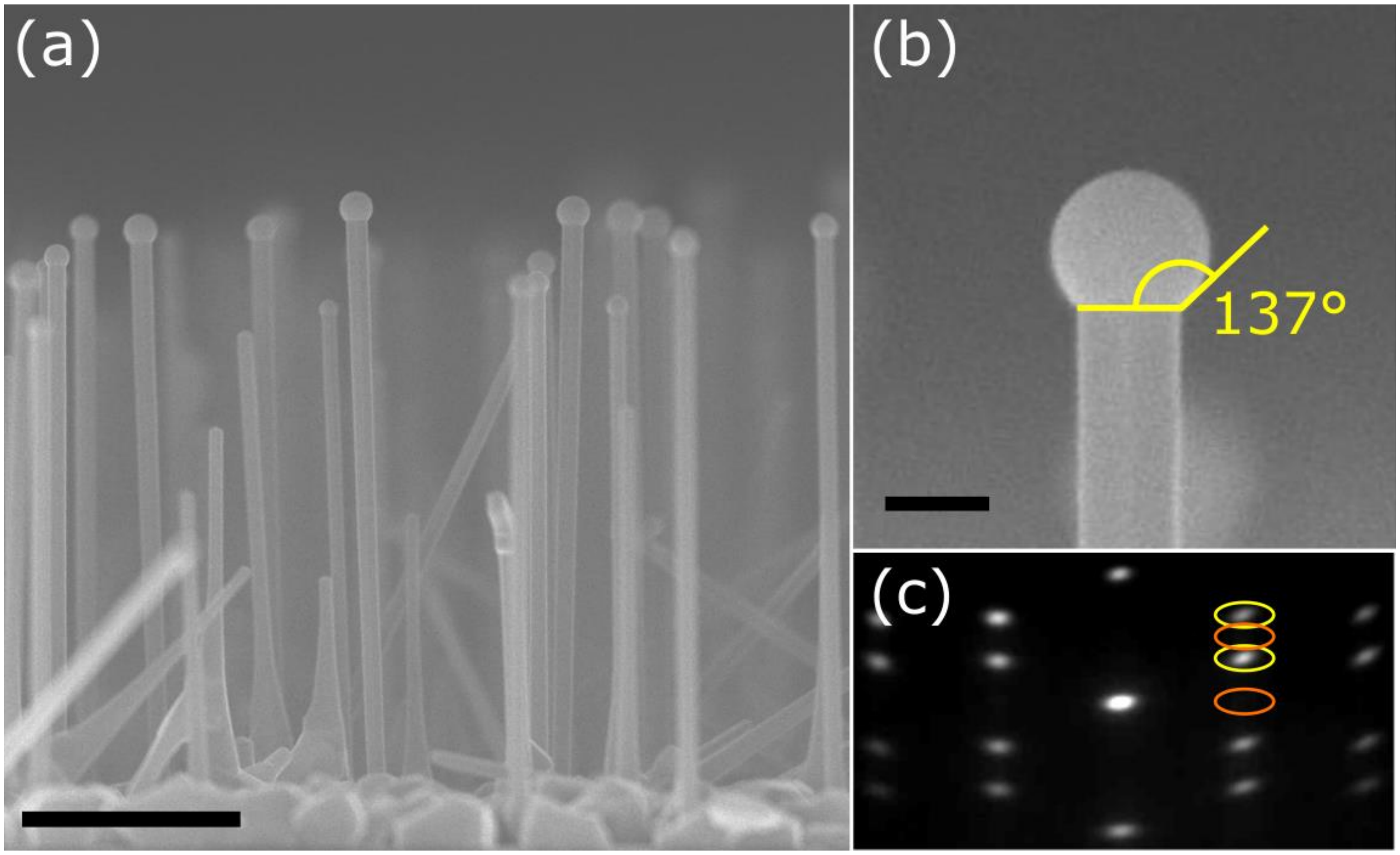

**Figure 2:** (a) and (b) SEM images showing the NW morphology after the standard growth of 25 min. Scale bars are 1 µm and 100 nm, respectively. (c) RHEED pattern recorded along the [1-10] azimuth at the end of the growth where only ZB spots (indicated by the yellow circles) are visible.

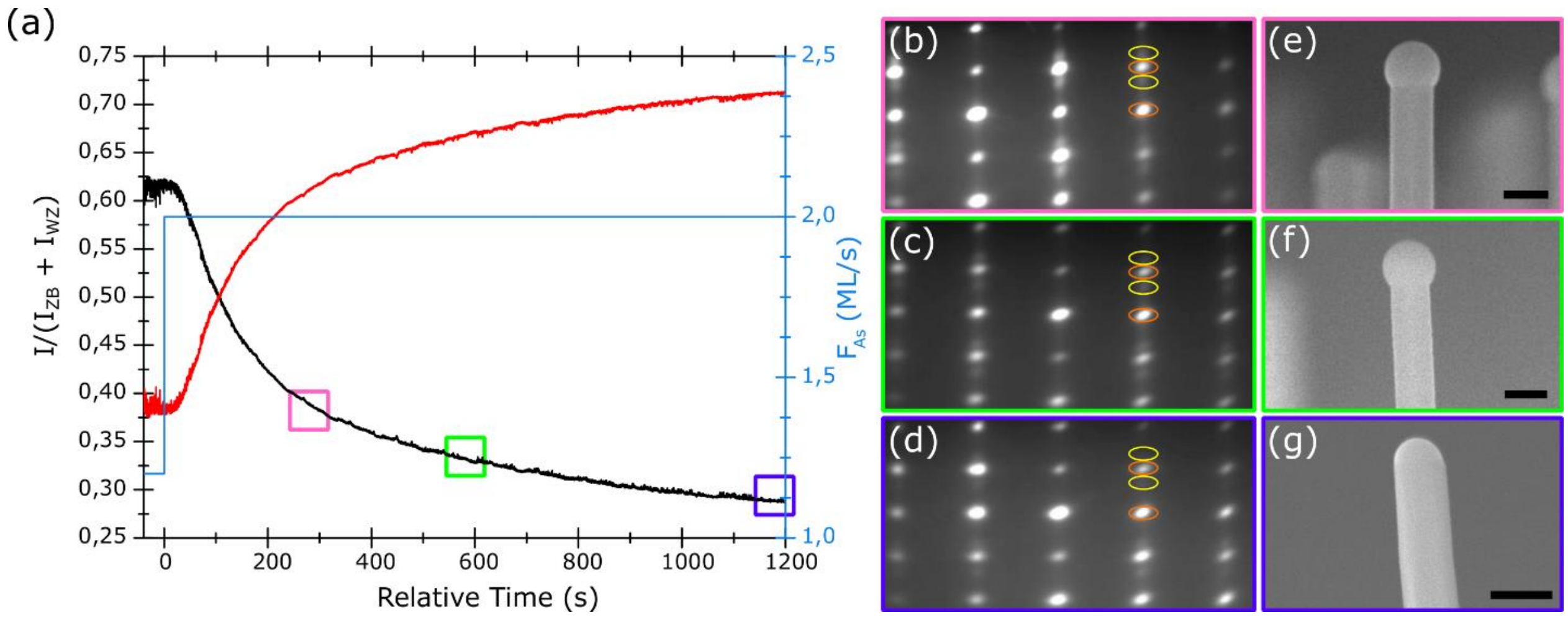

**Figure 3:** (a) $\frac{I_{ZB}}{I_{ZB}+I_{WZ}}$ and $\frac{I_{WZ}}{I_{ZB}+I_{WZ}}$ intensity ratios as a function of the relative growth time (t=0 s corresponds to the increase of the As flux, after the 25 min growth under the standard conditions). The light blue curve corresponds to the As flux in ML/s. On the IR horizontal axis, t=0 s corresponds to the increase of the As flux. The time indicates the growth time under a high As flux, following the 25 min growth under standard conditions. RHEED pattern recorded along the [1-10] azimuth and SEM images of the catalyst droplet after (b) and (e) 5 min, (c) and (f) 10 min, (d) and (g) 20 min of growth under a high As flux. The ZB and WZ spots are highlighted with the yellow and orange circles, respectively. Scale bars are 100 *nm*.

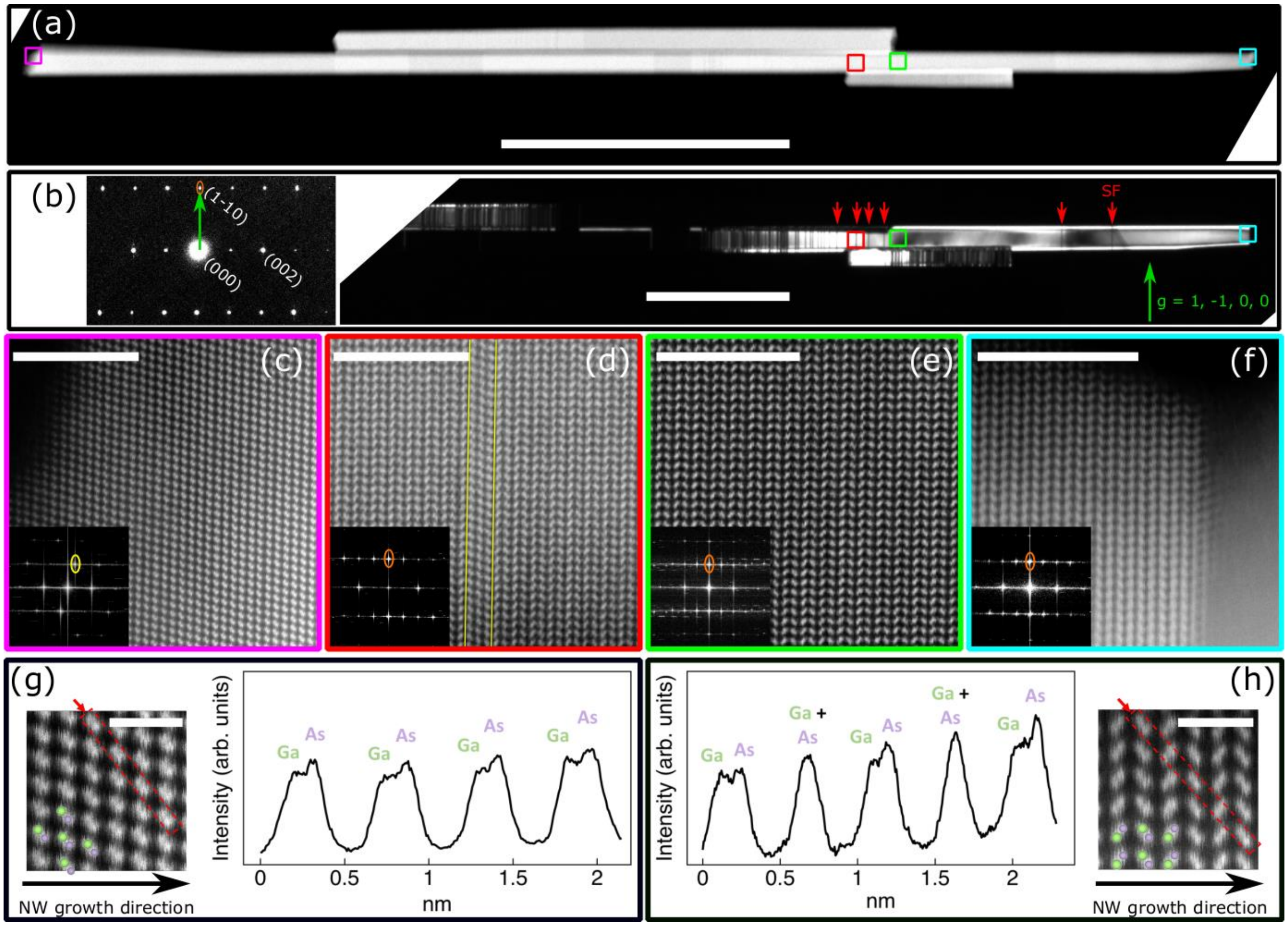

**Figure 4:** (a) STEM-HAADF overview of a single NW. Broken pieces of other NWs are positioned at the top and bottom. Scale bar: 1 µm. (b) DF-TEM image of the extended WZ segment. The red arrows indicate the stacking faults visible in the WZ segment. Scale bar: 500 nm. High-resolution STEM-HAADF images showing (c) the foot of the NW with a ZB crystalline structure, (d) the beginning of the extended WZ segment with a 3 MLs long ZB segment, (e) a portion of the extended WZ segment and (f) the end of the extended WZ segment with the Ga droplet/NW interface. (c, d, e, f) correspond to the regions highlighted with colored squares, magenta, red, green, and cyan, respectively, in (a) and (b). The ZB and WZ spots of the FFT are highlighted with the yellow and orange circles, respectively. Scale bars on (c) - (f) are 5 *nm*. The ZB structure in (c, d, g) is observed in the [110] zone axis, and the WZ structure in (e, f, h) is seen in the $[11\bar{2}0]$ zone axis. (g) High-resolution STEM-HAADF image and intensity profile of the ZB segment near the NW foot. (h) High-resolution STEM-HAADF image and intensity profile of the WZ segment near the NW head. Scale bars on (g) and (h) are 1 *nm*. The overlayed atomic models of WZ and ZB highlight Ga (green) and As (purple) atomic column positions, and were made using the VESTA software[54].

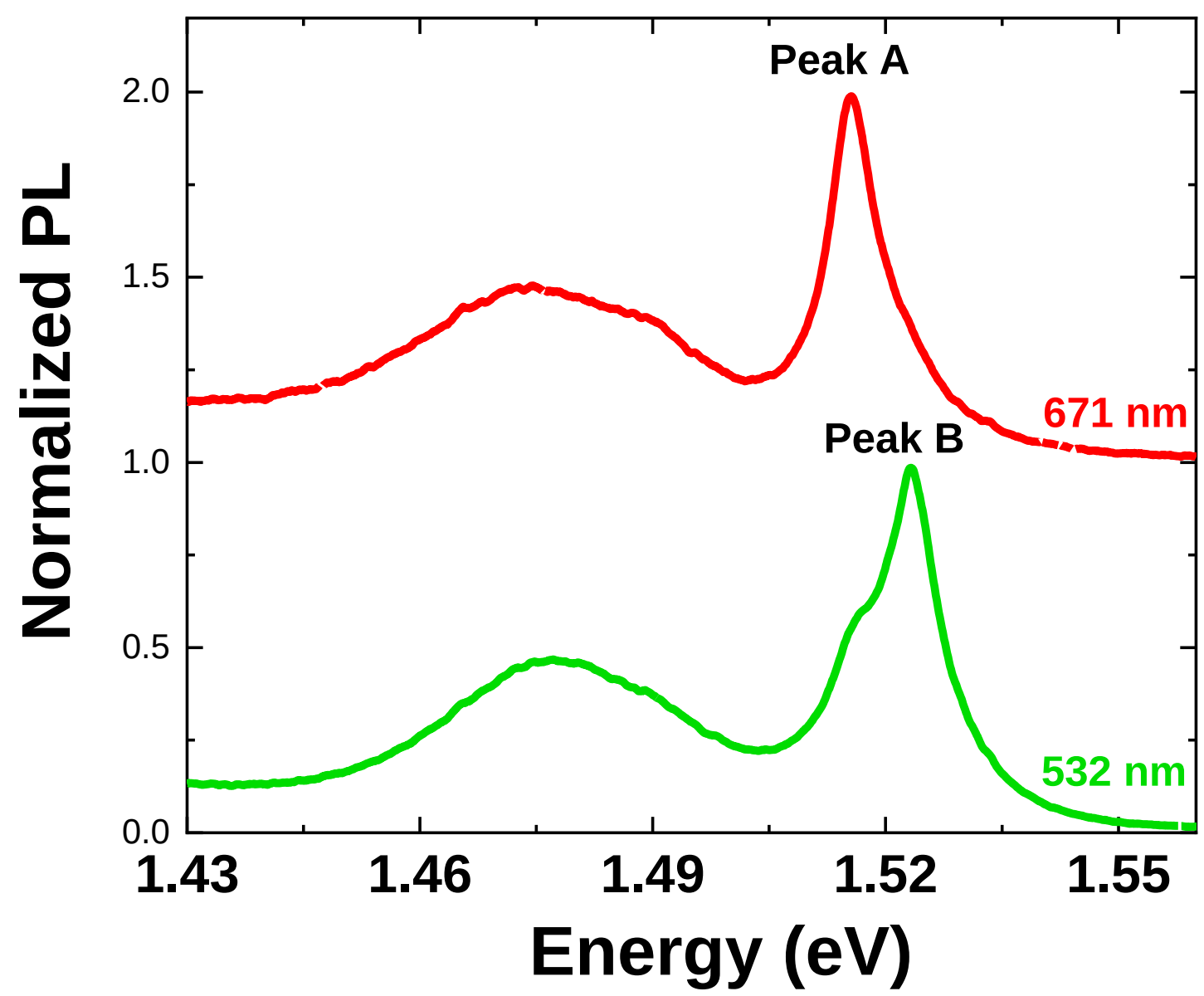

**Figure 5**: Photoluminescence spectra of the sample at 12 K for a 671 nm excitation wavelength (red curve) and a 532 nm excitation wavelength (green curve). The spectra have been normalized to the strongest emission peak and offset for clarity.

**Figure SI:**

**Table S1:** Measured and simulated values of the wetting angle $\beta$ for different growth conditions. The first row corresponds to a standard growth of the self-catalyzed GaAs NWs, with a 2.4 V/III flux ratio. Every other rows correspond to a standard growth, followed by an extended growth with a higher As flux.

| Growth time (minutes) | V/III flux ratio | Measured β (deg) | Simulated β (deg) |
|---|---|---|---|
| 25 | 2.4 | 137 | 140 |
| 25 + 20 | 2.4 – 3.0 | 135 | 135 |
| 25 + 20 | 2.4 – 3.7 | 127 | 125 |
| 25 + 10 | 2.4 – 4.0 | 120 | 120 |
| 25 + 5 | 2.4 – 5.1 | 107 | 110 |
| 25 + 5 | 2.4 – 5.8 | | 105 |

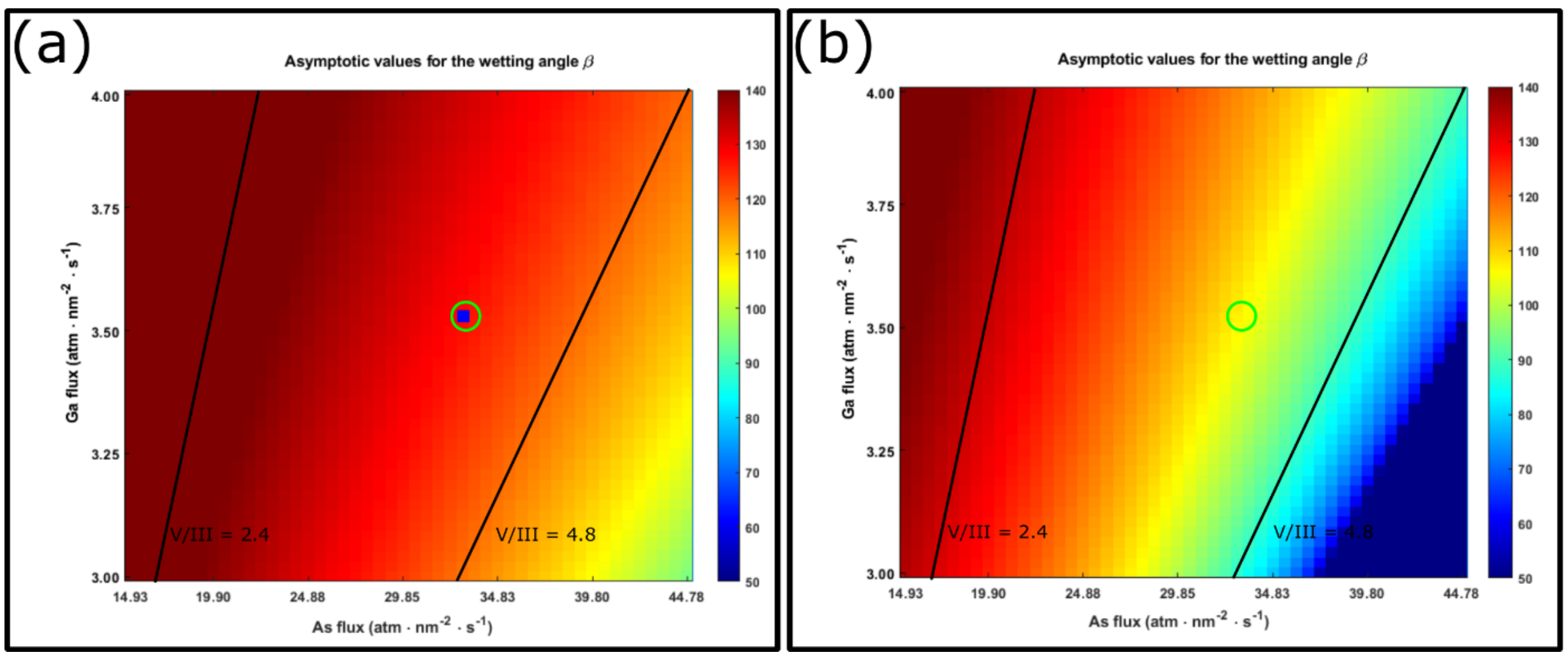

**Figure S1:** Values of the wetting angle $\beta$ obtained for a NW with (a) $r = 40\ nm$ and (b) $r = 60\ nm$ for various V/III flux ratio. Values lower than 50° are not represented as we consider that the droplet disappears and the VLS growth stops. The green circle represents the expected wetting angle obtained with our extended WZ growth conditions (V/III flux ratio = 4.0 at $q_{Ga} = 3.53\ atm/(nm^2.s)$). The experimental values of the wetting angle obtained on tens of NWs after an extended WZ growth of 20 min are represented with the blue square in (a).

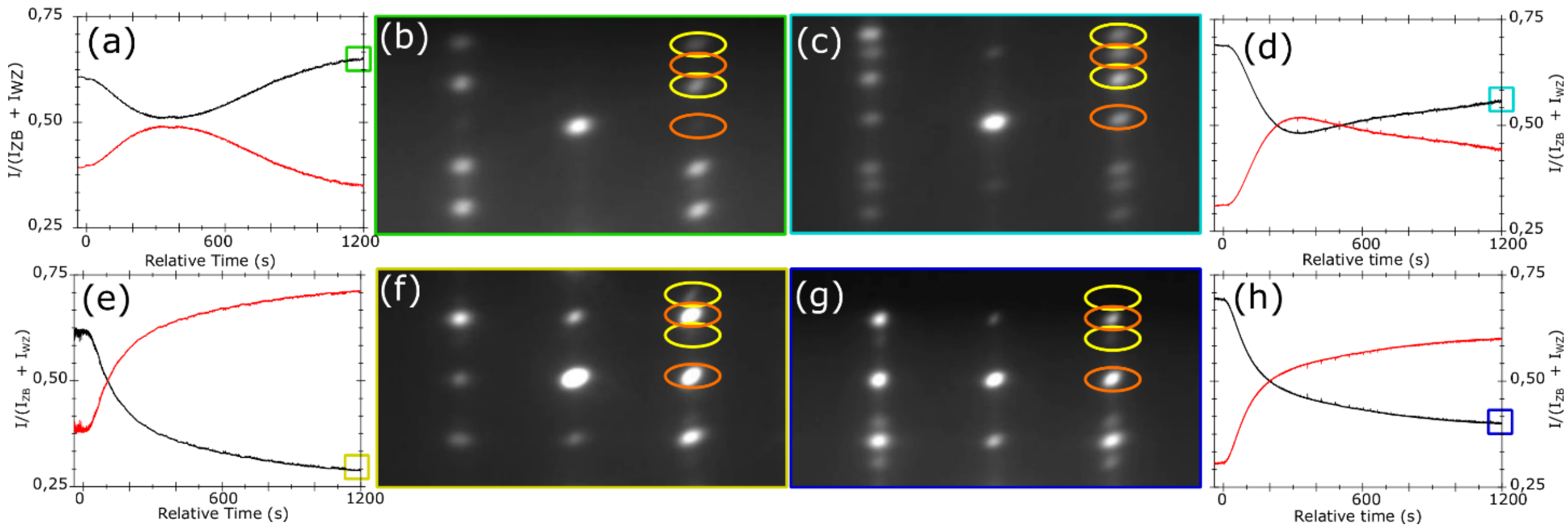

**Figure S2:** $\frac{I_{ZB}}{I_{ZB}+I_{WZ}}$ (black curves) and $\frac{I_{WZ}}{I_{ZB}+I_{WZ}}$ (red curves) intensity ratios obtained during the 20 min growth of the extended WZ segment and RHEED patterns at the end of the growth with a V/III flux ratio of (a) and (b) 3.0, (c) and (d) 3.5, (e) and (f) 4.0, (g) and (h) 4.3. On the IR horizontal axis, t=0 s corresponds to the increase of the As flux. The time indicates the growth time under a high As flux, following the 25 min growth under standard conditions. Yellow and orange circles indicate the ZB and WZ spots, respectively.

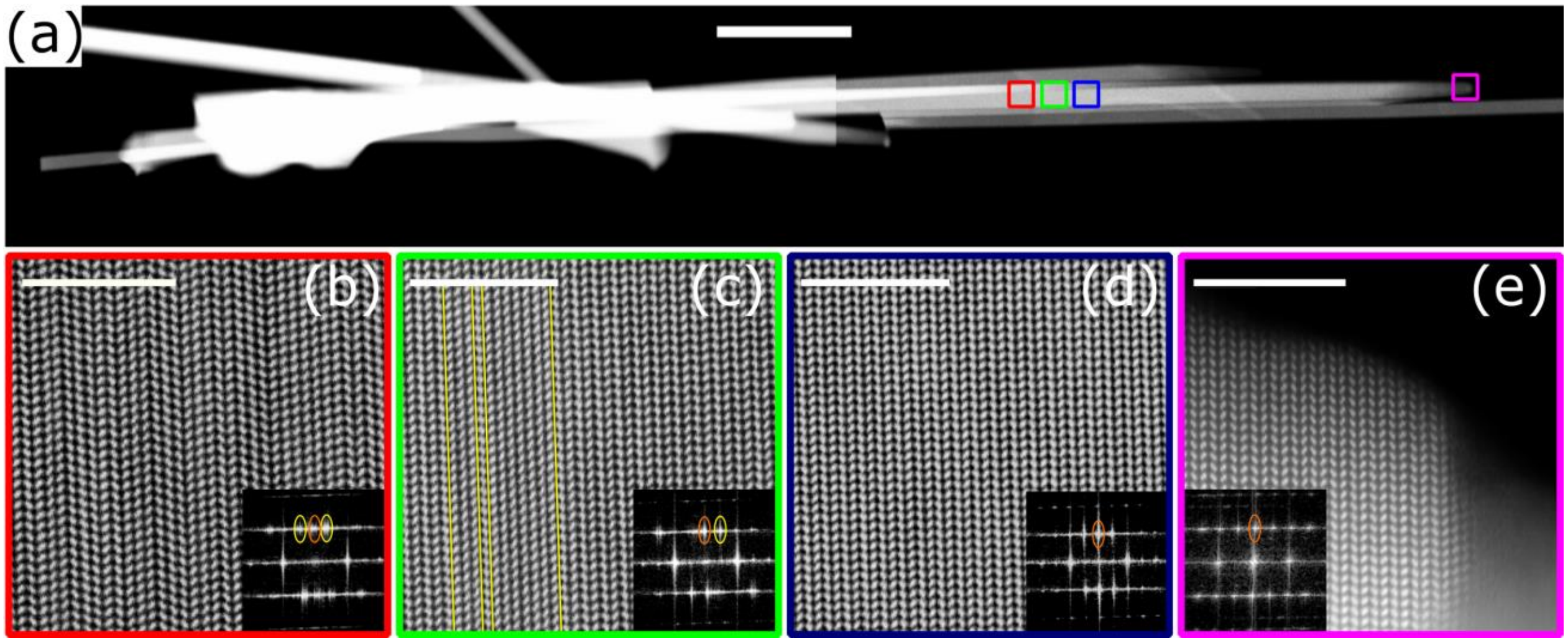

**Figure S3:** (a) STEM-HAADF overview of a NW cluster. Scale bar is 0.5 $\mu m$. High-resolution STEM-HAADF images showing (b) the defect section of the ZB – to – WZ transition, the beginning of the extended WZ segment with few MLs long ZB segments, (d) a portion of the extended WZ segment and (e) the end of the WZ segment with the Ga droplet/NW interface. (b, c, d, e) correspond to the regions highlighted with colored squares, red, green, blue and magenta in (a). The ZB structure in (b and c) is observed in the [110] zone axis, and the WZ structure in (c, d, e) is seen in the $[11\bar{2}0]$ zone axis. Scale bars on (b) – (e) are 5 $nm$.

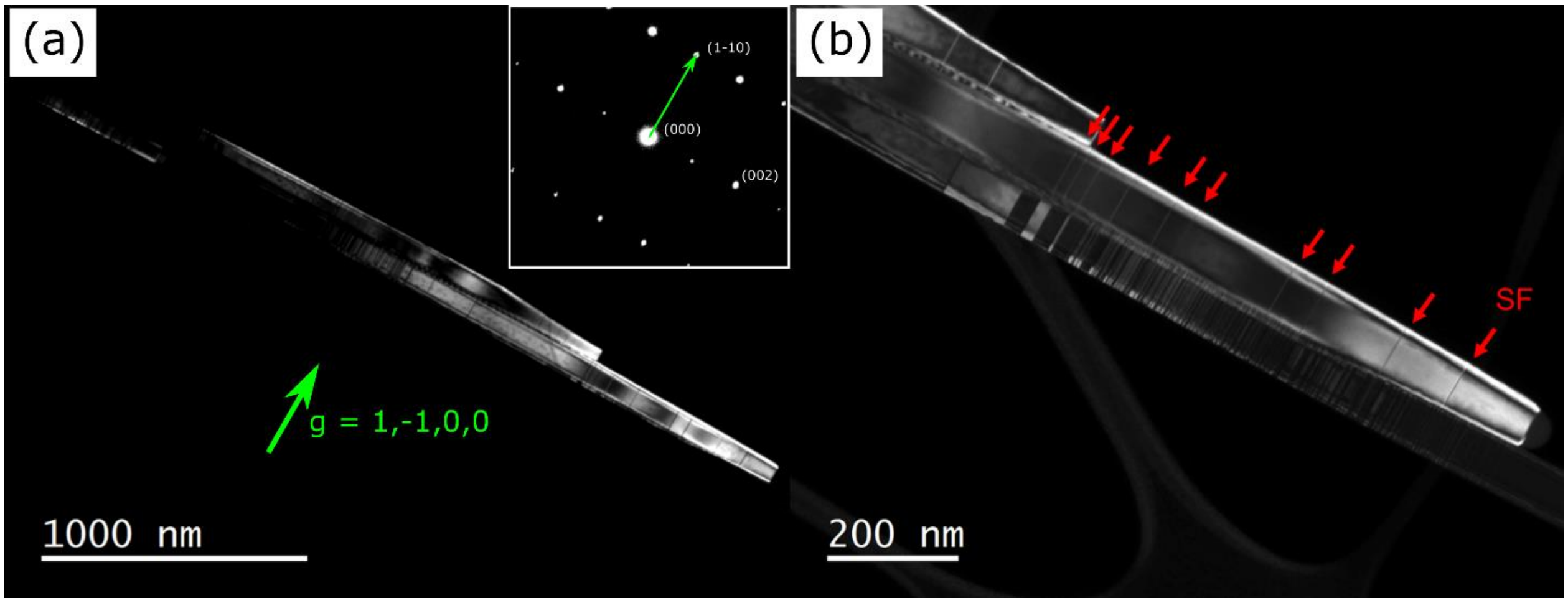

**Figure S4:** (a) DF-TEM image of the extended WZ segment. (b) Highlighting of few MLs long ZB insertions along the WZ segment.

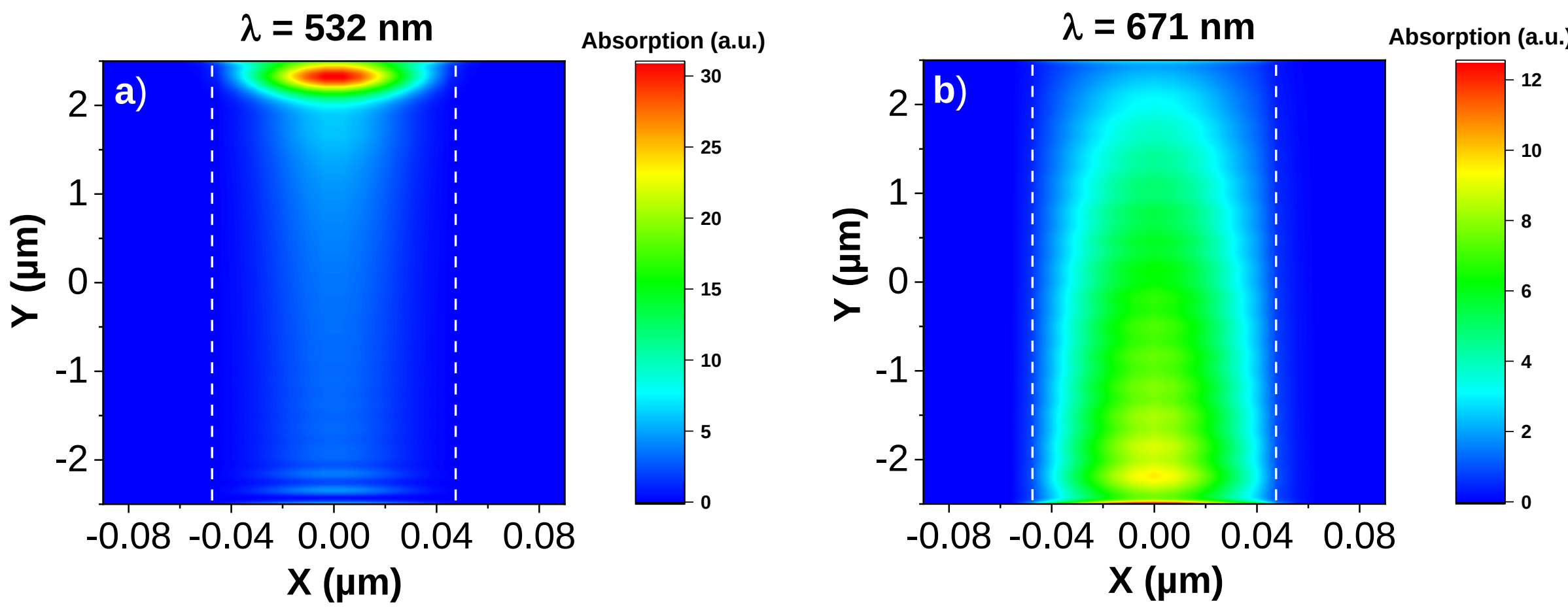

**Figure S5:** absorption profile for a 2D-cut along the nanowire growth axis for (a) λ=532 nm and (b) λ=671 nm plane waves.

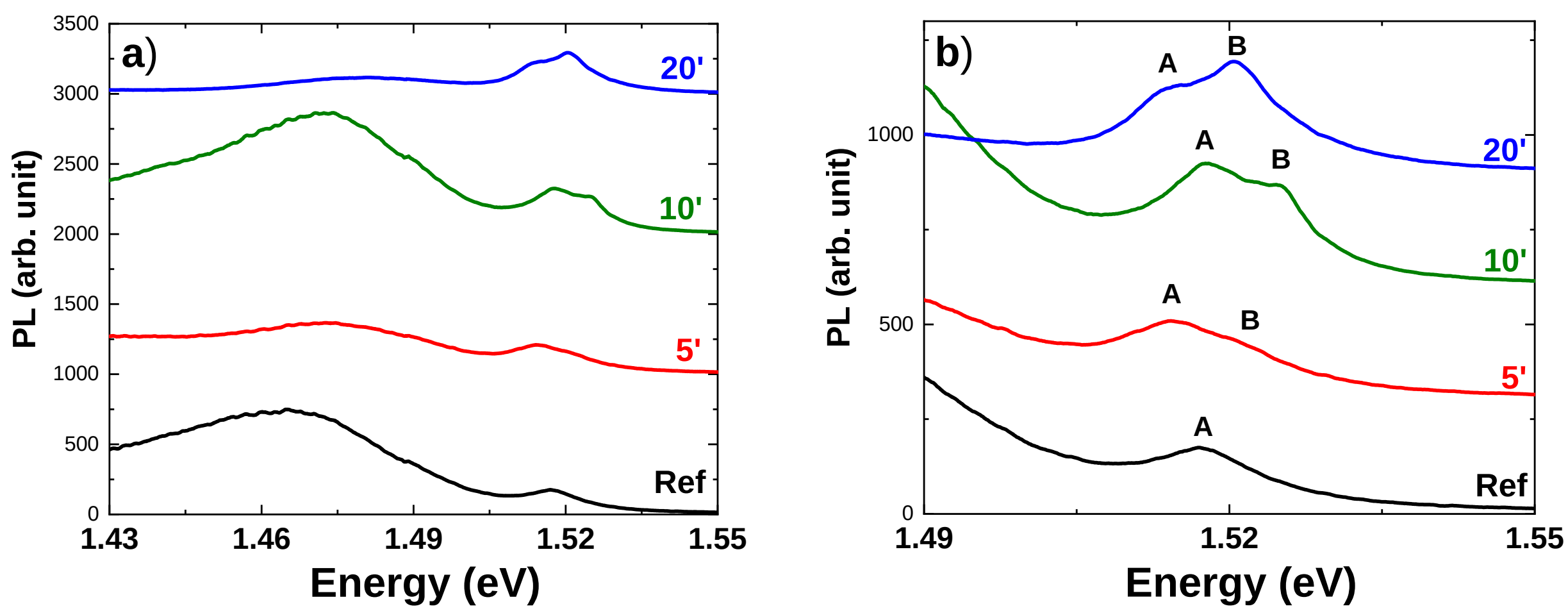

**Figure S6:** (a) Photoluminescence spectra of the samples at 12K for a 532 nm laser excitation (35 W/cm²). (b) Close-up view on the peaks related to free excitons. The spectra have been offset for clarity.

| | Growth time (min) | V/III flux ratio | Extended WZ growth time (min) | Extended WZ V/III flux ratio |
|---|---|---|---|---|
| Sample 1 (reference) | 25 | 2.4 | – | – |
| Sample 2 | 25 | 2.4 | 5 | 4.0 |
| Sample 3 | 25 | 2.4 | 10 | 4.0 |
| Sample 4 | 25 | 2.4 | 20 | 4.0 |

**Table S2:** Growth time of the different samples used in the PL characterization.